\documentclass[11pt,preprint2]{aastex}
\graphicspath{{./}{}}
\usepackage{float}
\usepackage[caption=false]{subfig}
\usepackage{amsmath}
\providecommand{\DIFdel}[1]{} 

\begin{document}
\title{The impact of applying black hole-host galaxy scaling relations to large galaxy populations}
\author[0000-0001-7829-4764]{Maggie C. Huber}
\author[0000-0003-1407-6607]{Joseph Simon}
\altaffiliation{NSF Astronomy \& Astrophysics Postdoctoral Fellow}
\author[0000-0001-8627-4907]{Julia M. Comerford}
\affiliation{University of Colorado Boulder, Boulder, CO 80309, USA}
\correspondingauthor{Maggie C. Huber}
\email{margaret.huber@colorado.edu}

\begin{abstract}
Supermassive black holes (SMBHs) with dynamically measured masses have shown empirical correlations with host galaxy properties. These correlations are often the only method available to estimate SMBH masses and gather statistics for large galaxy populations across a range of redshifts, even though the scaling relations themselves are derived from a small subset of nearby galaxies. Depending on the scaling relation used, estimated SMBH masses can vary significantly. The most widely used scaling relations are the M$_{BH}-$M$_{\mathrm{bulge}}$ and M$_{BH}- \sigma$ relations, where M$_{\mathrm{bulge}}$ is galaxy bulge mass and $\sigma$ is the bulge velocity dispersion. In this paper, we determine how severely the choice of scaling relation
impacts SMBH mass estimates for different subsets of a large galaxy population.
For this analysis we use a sample of $\sim$ 400,000 galaxies, including 1,240 Type 1 AGN from the Sloan Digital Sky Survey. We calculate SMBH masses from M$_{BH}-$M$_{\mathrm{bulge}}$ and M$_{BH}- \sigma$ and compare to single-epoch virial SMBH masses from broad-line H$\beta$, which are derived independently of black hole-host galaxy scaling relations. We find that SMBH masses derived from the single-epoch virial relation for H$\beta$ are better reproduced by M$_{BH}- \sigma$ than M$_{BH}-$M$_{\mathrm{bulge}}$. Finally, in cases where $\sigma$ and M$_{\mathrm{bulge}}$ cannot be measured directly, we show that it is possible to infer $\sigma$ from photometry with more accuracy than we can infer M$_{\mathrm{bulge}}$.
\end{abstract} 

\section{Introduction} \label{introduction}

Supermassive black holes (SMBHs) play a key role in galaxy formation and evolution, as almost all galaxies are assumed to harbor a SMBH at their center \citep[e.g.,][]{kormendyrichstone1995,Kormendy2013}. The question of how SMBHs grow their mass and co-evolve with their host galaxies is the subject of active research, and observational constraints on SMBH binaries can provide crucial insight into SMBH evolution. One promising potential observation of the SMBH binary population comes from the gravitational wave background (GWB) detected by pulsar timing arrays (PTAs) \citep[e.g.,][]{gwb,epta}. The GWB can constrain SMBH merger timescales and lead to a breakthrough in our knowledge of how galaxy mergers seed black hole mass growth. However, this observation hinges on our understanding of the SMBH binary population in our universe \citep{astrointerp}. 

More specifically, GWB analysis relies on an accurate SMBH mass function and merger rate to constrain evolutionary timescales \citep[e.g.,][]{Simon:2016,Simon:2023dyi}. The SMBH mass function and galaxy merger rate can be estimated using observations from galaxy surveys covering massive (M$_* > 10^{10}$M$_\odot$) systems out to redshift $z\sim3$. These galaxies are representative of the SMBH binary population that dominates the GWB \citep[e.g.,][]{Sesana:2013,Ravi:2015}. 

For most large galaxy samples, particularly at high redshifts, it is impossible to get dynamical measurements to directly measure SMBH masses. Therefore, observational constraints on the SMBH mass function often must be estimated indirectly via scaling relations and are hindered by uncertainty that increases with redshift \citep[e.g.,][]{vanderwel2014,Ferre-Mateu2015,cayenne2023,Comeron2023}. This uncertainty impacts all galaxy population statistics that use black hole-host galaxy scaling relations. For example, the choice of scaling relation can affect galaxy evolution models and cosmological simulations that rely on scaling relations as a benchmark for calibration \citep[e.g.,][]{eagle,tng,simba,Habouzit2021}. For many such analyses, scaling relations and indirectly measured host galaxy properties are the only tools at our disposal to determine the SMBH mass function.

 Black hole mass scaling relations are built on direct SMBH mass measurements of local galaxies. Direct measurements come from dynamical modeling of spatially resolved stars and gas with Keplerian orbits around the central black hole of a galaxy \citep[e.g.,][]{Ghez2005,Gultekin2009,McConnell2011}. The Event Horizon Telescope has also measured black hole mass directly for M87 and Sgr A* by determining the angular gravitational radius from geometric modeling or magneto-hydrodynamic simulations \citep{EHT2019,EHT2022}.  Dynamical modeling is limited to a small population of galaxies in the local universe that can be observed with high enough resolution to probe the central region of the host galaxy.
 
 To extrapolate these direct mass estimates out to black holes beyond the local universe, we take advantage of empirical correlations of dynamical SMBH mass with more easily observable host galaxy properties such as stellar mass and kinematics \citep[e.g.,][]{2000ApJ...539L...9F,2000ApJ...539L..13G,2002ApJ...574..740T,2003ApJ...589L..21M,2009ApJ...698..198G,Kormendy2013,McConnell:2013,remco2016,denicola2019}. M$_{BH} - $M$_{\mathrm{bulge}}$ and M$_{BH} - \sigma$ are widely used scaling relations that estimate SMBH mass using the host galaxy bulge mass and bulge velocity dispersion, respectively. The samples from which scaling relations are derived are inherently limited to $\sim30-100$ galaxies where it is possible to directly measure bulge mass, bulge velocity dispersion, and SMBH mass.  To ensure accurate SMBH mass estimates, the galaxy types and stellar masses used for scaling relations should be representative of the larger population to which the relations are applied. A major source of error when predicting SMBH masses is the bias in the galaxy sample used to fit scaling relations \citep[e.g.,][]{Kormendy2013,McConnell:2013,graham2016}.
 
Scaling relations limit their sample to certain galaxy types partly because of disagreement about whether black hole-host galaxy correlations are valid for all galaxies, regardless of morphology. The \cite{Kormendy2013} scaling relations do not include galaxies with pseudobulges, the dense central regions of disk galaxies with features similar to those of elliptical galaxies. \cite{Kormendy2013} argue that secular evolution affects pseudobulge properties with no correlation to SMBH mass. \cite{McConnell:2013} takes a different approach, distinguishing between late- and early-type galaxies. They use both galaxy types when constructing the M$_{BH} - \sigma$ relation, whereas M$_{BH} - $M$_{\mathrm{bulge}}$ is only derived from early-type galaxies. 
 
 More recent approaches use separate scaling relations for different galaxy morphologies. The galaxy morphology classification is based on detailed decompositions of each galaxy component that isolate the mass of the spheroid \citep[e.g.,][]{atlasscaling, scott2013, Davis2019,Graham2023,Graham2023b}. These studies cite that the discrepancies between M$_{BH} - $M$_{\mathrm{bulge}}$ and M$_{BH} - \sigma$ arise from a non-log-linear relationship between velocity dispersion and spheroid mass \citep[e.g.,][]{grahamdriver2007,graham2012}. This allows for M$_{BH} - $M$_{\mathrm{bulge}}$ relations for spiral galaxies that lack a classical bulge. To obtain the mass of the spheroid using morphology-dependent scaling relations, the photometry must be accurately decomposed into multiple components, including spheroids, disks, bars, and rings. The S$\mathrm{\acute{e}}$rsic profile and the mass-to-light ratios must be obtained individually for each component. Using this method to obtain a more accurate bulge mass decreases scatter in the M$_{BH} - $M$_{\mathrm{bulge}}$ relation, bringing it closer to the level of M$_{BH} - \sigma$ \citep{Davis2017}. For a small sample of nearby galaxies, morphology-dependent scaling relations are ideal and necessary for predicting the most accurate stellar mass-based SMBH masses. 

 However, morphology-dependent scaling relations have not yet been applied to large samples of galaxies with higher redshift (z $\gtrsim0.1$), due to the difficulty of accurately resolving detailed substructures \citep{s11}. Furthermore, two-component bulge+disk decompositions are not available for most galaxies with z $\gtrsim 0.1$. Therefore, the bulge mass and morphology for these galaxies are less accurate, preventing the correct usage of morphology-dependent scaling relations.

Additionally, as the light from most higher-redshift galaxies cannot be decomposed and many of these galaxies lack spectroscopic measurements, their SMBH mass estimates from M$_{BH} - $M$_{\mathrm{bulge}}$ and M$_{BH} - \sigma$ require inferred M$_{\mathrm{bulge}}$ and $\sigma$. In practice, M$_{\mathrm{bulge}}$ is inferred by prescribing a bulge fraction, $f_{\mathrm{bulge}}$, and multiplying by the total stellar mass of the galaxy \citep[e.g.,][]{Sesana:2013,Ravi:2015,Arzoumanian:2021}. The $f_{\mathrm{bulge}}$ prescription is often based on galaxy properties that are approximately related to the actual bulge fraction, such as galaxy color and total stellar mass. The velocity dispersion, $\sigma$, can be inferred from the total stellar mass of the galaxy, the effective radius, and the S$\mathrm{\acute{e}}$rsic index by using the virial theorem with a S$\mathrm{\acute{e}}$rsic-dependent virial constant \citep[e.g.,][]{Bezanson:2011}. For higher-redshift galaxies where the S$\mathrm{\acute{e}}$rsic index is not available, one can also infer velocity dispersion from the mass fundamental plane (MFP) using the effective radius and total stellar mass \citep{mfp}.
 
 In addition to the need to infer host galaxy properties, redshift evolution complicates SMBH mass estimation for high-z galaxies. Whether M$_{BH} - $M$_{\mathrm{bulge}}$ and M$_{BH} - \sigma$ evolve with redshift is still unclear \citep[e.g.,][]{croton2006,robertson2006,woo2006,Woo2008,merloni2010,vanderwel2014,shen2015b,Habouzit2021,mountrichas2023}. Even assuming scaling relations themselves do not evolve with redshift, redshift evolution in the galaxy mass-size relationship arises because higher-redshift galaxies tend to be more compact \citep[e.g.,][]{vanderwel2014}. More compact galaxies have smaller effective radii for a fixed stellar mass. When we use stellar mass and effective radius to infer $\sigma$, the SMBH masses from the M$_{BH} - \sigma$ relation capture evolution in the mass-size relationship that is absent when using only stellar mass to estimate M$_{\mathrm{BH}}$. For galaxies at $ 1 < z < 3$, this leads to an inferred higher-redshift SMBH population from M$_{BH} - \sigma$ that diverges from the population inferred by M$_{BH} - $M$_{*}$ \citep{cayenne2023}.

The disagreement between scaling relations and issues with their application are widely known \citep[e.g.,][]{grahamdriver2007,graham2016,vandenbosch2016,cayenne2023,Sturm2024}, but they are still often the only method that allows us to estimate SMBH mass on a galaxy population scale. In this paper, we outline the severity of these issues by comparing the SMBH mass estimates from different scaling relations for a sample of $\sim400,000$ galaxies from the Sloan Digital Sky Survey (SDSS) between redshift $0.02 < z < 0.2$ with a well-defined M$_{\mathrm{bulge}}$ and $\sigma$. Furthermore, we calculate the SMBH mass for $1,240$ Type 1 AGN in this SDSS sample using a new single-epoch virial mass estimator from \cite{Shen2024} that is independent of scaling relations. We compare the virial SMBH mass estimates with SMBH mass estimates from M$_{BH} - $M$_{\mathrm{bulge}}$ and M$_{BH} - \sigma$. We also use our well-defined M$_{\mathrm{bulge}}$ and $\sigma$ to assess the validity of inferring these properties without decomposed photometry and spectroscopic data, which is necessary for a wide range of astrophysical studies that require SMBH masses for higher-redshift galaxies.

This paper is organized as follows: Section \ref{methods} discusses how we constructed our galaxy sample, obtained measured and inferred host galaxy properties, and the scaling relations we use. Section \ref{results} shows our results comparing M$_{BH} - $M$_{\mathrm{bulge}}$ and M$_{BH} - \sigma$, inferred and measured galaxy properties, and SMBH masses estimated from scaling relations and single-epoch virial estimates for Type 1 AGN. Section \ref{conclusion} summarizes our findings.

\section{Data} \label{methods}

The dataset used for this project comes from SDSS DR7 \citep{sdssdr7}. DR7 is the largest SDSS data release with available bulge-disk decompositions performed by \cite{mt14}, pure S$\mathrm{\acute{e}}$rsic decompositions by \cite{s11}, and broad-line spectral detections from \cite{Liu:2019}. To construct a galaxy sample from these catalogs, we start with the 657,996 galaxies in \cite{mt14} and follow the redshift and mass cuts described in \cite{Thanjavur:2016}, restricting the catalog to $0.02 \leq z \leq 0.2$ and $9 \leq \log(M_*/M_\odot) \leq 12$ for mass completeness. For each galaxy, \cite{mt14} reports total mass from both the one and two-component fits to the surface brightness profile. The one-component total mass uses the mass inferred by fitting the entire galaxy, and the two-component fit uses the separate bulge and disk component masses added together. Following the advice from \cite{mt14}, we exclude galaxies with disagreement between the total stellar mass from the one-component fit, M$_\mathrm{bulge+disk}$, and the mass from the two-component fit, M$_\mathrm{bulge}+$M$_\mathrm{disk}$. The catalog has the offset between the total stellar masses as $\Delta_{\mathrm{bulge+disk}}$ in units of standard error. We restrict our sample to galaxies where $\Delta_{\mathrm{bulge+disk}} < 1\sigma$. This should remove unreliable masses from our sample.

To break down the sample according to morphology, we compute a bulge-to-total mass fraction ($B/T$) using the sum of the bulge and disk component masses from \cite{mt14} for total stellar mass. Galaxies with a high S$\mathrm{\acute{e}}$rsic index are often incorrectly classified as two-component galaxies when a portion of the surface brightness profile is steeper than a deVaucouleurs profile \citep{Bluck:2014}. Following the results of `false' disks in simulated mock galaxies fit with the GIM2D software from \cite{Bluck:2014}, we correct galaxies where the uncorrected $B/T > 0.7$ and where the probability ($P_{pS}$) for a single component pure S$\mathrm{\acute{e}}$rsic morphology from \cite{s11} is greater than $0.32$ and the S$\mathrm{\acute{e}}$rsic index $n > 4$. For galaxies with a false disk, we set $B/T = 1$ and the morphology type to single-component elliptical (Type 1 in \citealt{mt14}).

\begin{figure}
    \centering
    \includegraphics[width=0.5\textwidth]{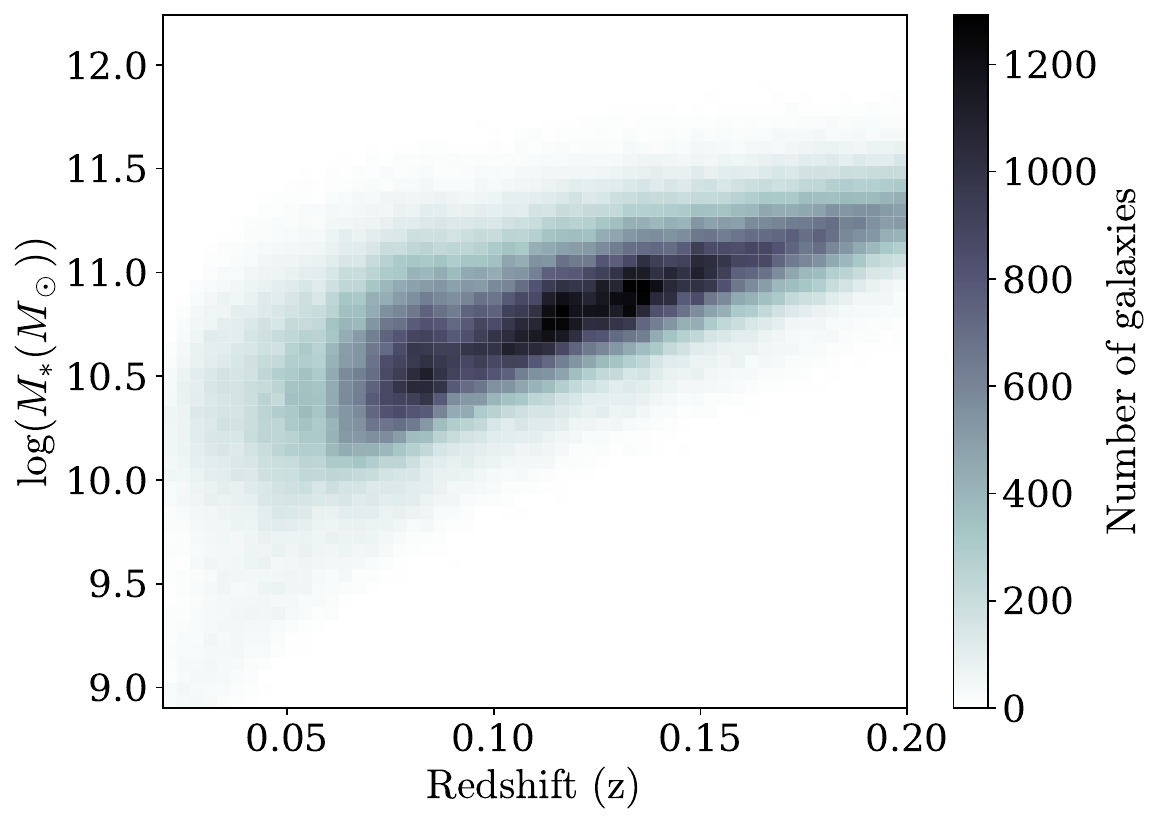}
    \caption{Here, our sample of 413,538 galaxies is visualized as a 2D histogram on a total mass vs. redshift plot. The colorbar indicates the galaxy count in each bin. \label{fig:totalmassredshift}}
\end{figure}

For velocity dispersion, we follow SDSS suggestions to only use velocity dispersions within the instrumental resolution of SDSS spectra where $70 < \sigma_{\mathrm{ap}}(\mathrm{km \ s}^{-1}) < 420$ for the uncorrected velocity dispersion measured in the 3$^{\prime\prime}$ SDSS aperture, $\sigma_{\mathrm{ap}} $. After making these cuts, we use 413,538 galaxies with well-defined measurements of M$_{\mathrm{bulge}}$ and $\sigma$ in our analysis. This sample of galaxies can be visualized using the total mass vs. redshift plot in Figure \ref{fig:totalmassredshift}.

To understand the divergence between scaling relations for galaxy types, we bin our sample by both total mass and morphology. For the total mass, we take the sum of the bulge and disk components from \cite{mt14} and choose four bins of $\sim103,000$ galaxies each. We draw these bins at $\log(M_*/M_\odot) = 9, 10.5, 10.8, 11, $ and $12$. 

We also bin the sample according to the galaxy type classification scheme from \cite{mt14} based on galaxy surface brightness profiles. Type 1 galaxies are best fit by a single de Vaucouleurs profile at all radii, with the distribution of S$\mathrm{\acute{e}}$rsic indices peaking at $n\sim4-5$. This accounts for most single-component, diskless elliptical galaxies. Type 2 galaxies are dominated by an exponential profile at all radii, with the distribution of S$\mathrm{\acute{e}}$rsic indices peaking at $n\sim1$. Type 2 includes both disk galaxies with a weak bulge component and galaxies that lack a bulge component in their S$\mathrm{\acute{e}}$rsic profile. In our sample, we only include Type 2 galaxies with weak bulges that have a measured M$_{\mathrm{bulge}}$ in order to test M$_{BH} - $M$_{\mathrm{bulge}}$. Type 3 galaxies are dominated by a de Vaucouleurs profile at central radii and an exponential profile at larger radii. These are either galaxies with both a disk and bulge component, face-on disks with a spurious bulge component, or extended single-component ellipticals with a spurious disk component. Type 4 galaxies include everything else that does not fit in Types 1-3; some examples are galaxies where the de Vaucouleurs and exponential components cross twice or if the exponential profile dominates in the center and de Vaucouleurs dominates at larger radii. Table \ref{tab:numgalaxies} shows the total number of galaxies in our total stellar mass and galaxy type bins. We attempt to correct any galaxy type misclassifications following the guidance in \cite{mt14} by correcting for false disks and limiting our sample to $\Delta_{\mathrm{bulge+disk}} < 1\sigma$. 

Our sample generally agrees with the results of morphology classification for local galaxies \citep[e.g.,][]{atlas3d2011,atlas3d2013}, with a difference in the relative abundance of diskless galaxies. \cite{atlas3d2013} found that 17$\%$ of their galaxy sample (31 of 180 galaxies) had no disk component within a total mass range of $9 < \log(\mathrm{M}_*/\mathrm{M}_\odot) < 12$, using stellar kinematics and photometry for 180 galaxies within a local volume of radius of $D=42$ Mpc. Our sample is only slightly different; 27$\%$ of (112,410 of 413,538 galaxies) our sample is Type 1 within $9 < \log(\mathrm{M}_*/\mathrm{M}_\odot) < 12$. The Type 1 classification is only based on the best-fitting S$\mathrm{\acute{e}}$rsic profile, with no stellar kinematic information available. Overall, the majority of our galaxy sample ($\sim70\%$ from Table \ref{tab:numgalaxies}) were found to have an exponential component in their S$\mathrm{\acute{e}}$rsic profile.

\begin{deluxetable*}{ccccc|c}
\tablecaption{Number of galaxies in total-mass + morphology categories
 \label{tab:numgalaxies}}
\tablehead{\colhead{Total mass range} & \colhead{Type 1 (bulge)}& \colhead{Type 2 (disk)}& \colhead{Type 3 (bulge+disk)}& \colhead{Type 4 (irregular)}& \colhead{Total}}
\startdata
$9 \leq \log(M_*) < 10.5 $   & 15,565 (15.2$\%$)  & 16,811 (16.4$\%$) & 65,279 (63.6$\%$)  & 4,974  (4.9$\%$) & 102,629 \\
$10.5 \leq \log(M_*) < 10.8$ & 23,627 (22.8$\%$)  & 9,308 (9.0$\%$)  & 67,246 (65.0$\%$)  & 3,347 (3.2$\%$) & 103,528 \\
$10.8 \leq \log(M_*) < 11$   & 31,303 (30.2$\%$)  & 4,877 (4.7$\%$) & 64,693 (62.3$\%$) & 2,960 (2.9$\%$) & 103,833 \\
$11 \leq \log(M_*) \leq 12$  & 41,915 (40.5$\%$)  & 1,425 (1.4$\%$) & 57,176 (55.2$\%$)  & 3,032 (2.9$\%$) & 103,548 \\
\hline
$9 \leq \log(M_*) \leq 12$  & 112,410 (27.2$\%$) & 32,421 (7.8$\%$) & 254,394 (61.5$\%$) & 14,313 (3.5$\%$) & 413,538
\enddata
\tablecomments{In parentheses is the percentage of galaxies of a specific type within the given total stellar mass range of each row.}
\end{deluxetable*}

\subsection{Measured bulge mass from bulge-disk decomposition}
Our galaxy sample uses the M$_{\mathrm{bulge}}$ from \cite{mt14} using SDSS photometry from the \cite{s11} catalog. \cite{s11} performs a Point-Spead-Function convolved bulge-disk decomposition on SDSS images using the GIM2D software with various S$\mathrm{\acute{e}}$rsic models, providing photometry for separate galaxy components. Two-component S$\mathrm{\acute{e}}$rsic models have limitations, as they do not account for complex substructures like galaxy bars and non-exponential disks \citep{2005disks,2005bars}. However, \cite{s11} note that they deliberately choose two-component fits to allow interfacing between their results and higher-redshift galaxy populations, where three-component fits are largely unavailable. This is especially important if we want to apply this study to interpretations of the GWB, since galaxies out to redshift $z \sim3$ contribute to the background. 

To convert from photometry to stellar mass, \cite{mt14} compare each galaxy's spectral energy distribution to various stellar population synthesis models with Bayesian analysis. This analysis gives the likelihood of the relevant stellar population parameters given the observed SED from \cite{s11}. The stellar mass is then determined from simulating a stellar population with the parameters given by the Bayesian analysis. 

It is important to note that $M_{\mathrm{*}}$ and $M_{\mathrm{bulge}}$ measurements are sensitive to assumptions about the initial mass function (IMF). This issue is most apparent on the high-mass end of the galaxy population \citep[e.g.,][]{Bernardi:2017,Liepold2024}, and potentially impacts our results from M$_{BH} - $M$_{\mathrm{bulge}}$. It could also impact inferred velocity dispersion, which uses total stellar mass, but also incorporates size and S$\mathrm{\acute{e}}$rsic index. The choice of IMF does not affect our spectroscopic velocity dispersion results. \cite{mt14} uses the \cite{chabrier} IMF and the \cite{calzetti} extinction law.   

Errors are reported in \cite{mt14} as the 16th and 84th percentile of the total mass PDF, which we convert to a standard deviation by taking the difference between these percentiles and the mean value. These values correspond to the upper and lower error bars we use when propagating bulge mass measurement uncertainty through black hole mass calculations.

For our analysis, we refer to the bulge mass calculated by \cite{mt14} as the measured M$_{\mathrm{bulge}}$, since it is the most direct way of calculating bulge mass for our galaxy sample.

\subsection{Inferred bulge mass from total stellar mass}
Although our galaxy sample has available bulge-disk decompositions, this is rarely the case for higher redshift galaxy samples due to resolution limits. Studies that require bulge mass measurements of a galaxy sample without decomposed photometry opt to infer M$_{\mathrm{bulge}}$. This is generally done by prescribing a bulge fraction using properties of the host galaxy to determine whether galaxies are more likely to be disk- or bulge-dominated.

Most bulge fraction estimates assign different bulge fractions to early (red) and late-type (blue) galaxies. To distinguish between early and late types, we adopt the color cut from \cite{Bluck:2014} where a galaxy is red if the global $g-r$ color is
\begin{equation}
\label{eq1}
    (g-r) > 0.06 \times \log(M_*/M_{\odot}) - 0.01,
\end{equation}
where $M_*/M_{\odot}$ is the total galaxy mass in solar masses.

The bulge fraction can then be assigned based on this color cut alone. For example, both \cite{Arzoumanian:2021} and \cite{Ravi:2015} use $f_{\mathrm{bulge}} = 1$ for early-types, while \cite{Arzoumanian:2021} uses $f_{\mathrm{bulge}} = 0.31$ and \cite{Ravi:2015} uses $f_{\mathrm{bulge}} = 0.2$ for late-types. \cite{Sesana:2013} uses a more complex $f_{\mathrm{bulge}}$ prescription, where it is randomly selected from the interval $[\langle f_{\mathrm{bulge}}\rangle - 0.1,\langle f_{\mathrm{bulge}}\rangle + 0.1]$ and $\langle f_{\mathrm{bulge}}\rangle$ is determined by the color cut and total mass. \cite{Sesana:2013} sets the average $\langle f_{\mathrm{bulge}}\rangle = 0.9$ for early-type galaxies with $M_* \geq  10^{11} M_\odot$, logarithmically decreasing to $\langle f_{\mathrm{bulge}}\rangle = 0.25$ for early-type galaxies at $M_* =  10^{10} M_\odot$.  For all late-type galaxies, $\langle f_{\mathrm{bulge}}\rangle = 0.2$.

In this paper, we propose an improved $f_{\mathrm{bulge}}$ prescription that uses results from kinematic M$_{\mathrm{bulge}}$ measurements \citep{c2011,c2013} that show decreasing $f_{\mathrm{bulge}}$ for late-type galaxies and early-type galaxies with $M_* <  10^{10} M_\odot$. We use the same random selection about the interval $[\langle f_{\mathrm{bulge}}\rangle - 0.1,\langle f_{\mathrm{bulge}}\rangle + 0.1]$, with $\langle f_{\mathrm{bulge}}\rangle = 0.9$ for early-type galaxies with $M_* \geq  10^{11} M_\odot$, logarithmically decreasing to $\langle f_{\mathrm{bulge}}\rangle = 0.10$ for early-type galaxies at $M_* =  10^{9} M_\odot$. For late-type galaxies with $M_* \geq  10^{11} M_\odot$, $\langle f_{\mathrm{bulge}}\rangle = 0.25$ logarithmically decreasing to $\langle f_{\mathrm{bulge}}\rangle = 0.10$ for late-type galaxies at $M_* =  10^{9} M_\odot$. We find that this prescription best recovers inferred M$_{\mathrm{bulge}}$ compared to the \cite{mt14} results, although there are still issues inherent in prescribing $f_{\mathrm{bulge}}$ that we discuss in Section \ref{ivm}.

\subsection{Measured velocity dispersion from spectra}
We obtain velocity dispersion measurements for our sample from the SDSS-DR18 spectroscopy catalog \citep{dr18}. SDSS measures velocity dispersion by fitting the spectrum to a catalog of eigenspectra templates broadened by Gaussian kernels ranging from 100 km s$^{-1}$ to 850 km s$^{-1}$, iterating by steps of 25 km s$^{-1}$. These templates are redshifted to the best-fit galaxy redshift, and the best-fit velocity dispersion is selected from the minimum chi-squared value. SDSS warns against using velocity dispersions with low signal-to-noise (S/N $<$ 10), which typically encompasses velocity dispersions outside the range of $70 < \sigma_{\mathrm{ap}}(\mathrm{km \ s}^{-1}) < 420$. We limit our sample accordingly as described in Section \ref{methods}.

The velocity dispersion from SDSS reflects the second moment of stellar velocity within the fixed aperture size and must be corrected to each galaxy's effective radius. For our sample, we use the same aperture correction as \cite{Bezanson:2011} where

\begin{equation} \label{apcorr}
    \sigma = \sigma_{ap}(8.0r_{ap}/r_e)^{0.066}.
\end{equation}

Here, the aperture velocity dispersion reported by SDSS is $\sigma_{ap}$ and $r_{ap} = 1.5 \arcsec $ is the SDSS spectroscopic fiber radius. Effective radius $r_e$ is given by 
\begin{align}
    r_e = R_{hl}\sqrt{1-e}
\end{align}
Where $R_{hl}$ is the half-light galaxy semi-major axis and $e$ is galaxy ellipticity, both from the \cite{s11} photometry catalog.

\subsection{Inferred velocity dispersion from total mass, size, and Sersic index} \label{siginf}
The availability of spectroscopic survey data, including spectroscopic velocity dispersion, is inherently limited. Large galaxy samples used to represent the universal population SMBHs primarily use photometric survey data. Fortunately, like bulge mass, velocity dispersion can also be estimated with photometric quantities. An elliptical galaxy's dynamical mass ($M_{dyn}$) is related to its velocity dispersion and size through the virial theorem. \cite{Taylor2010} provides a structure-corrected conversion between stellar and dynamical mass to extend this relation to all galaxy types. This allows us to infer dynamical mass from total mass with a dependence on the S$\mathrm{\acute{e}}$rsic index. 

We adopt the inferred velocity dispersion  \citep{Bezanson:2011} with a S$\mathrm{\acute{e}}$rsic dependent virial constant $K_{\nu}(n)$ from \cite{Bertin2002} where

\begin{equation}
    K_\nu(n) = \frac{73.32}{10.465 + (n-0.94)^2} + 0.954,
    \label{eq2}
\end{equation}
and the galaxy S$\mathrm{\acute{e}}$rsic index $n$ is from the \cite{s11} pure S$\mathrm{\acute{e}}$rsic decompositions.

We then use $K_{\nu}(n)$, $M_\star$, and effective radius $r_e$ to determine inferred velocity dispersion given in \cite{Bezanson:2011} with
\begin{equation}
    \sigma_{inf} = \sqrt{\frac{GM_{*}}{0.557K_{\nu}(n)r_e}}.
\label{eq3}
\end{equation}

Although Equation \ref{eq3} does not take into account rotational velocities of spiral galaxies, the structure-dependent virial constant in Equation \ref{eq2} is enough to align spiral and elliptical galaxies on the same stellar mass-effecitve radius-velocity dispersion fundamental plane \citep{Taylor2010,Bezanson:2011,deGraaff2021}. We find that velocity dispersions are slightly underestimated with Equation \ref{eq3} compared to those measured by spectroscopy for Type 2 galaxies. Measured $\sigma$ is generally higher than inferred $\sigma$ by a median factor of $0.1$ dex for Type 2 galaxies in our sample.

\subsection{M$_{BH}-$M$_{\mathrm{bulge}}$ and M$_{BH}-\sigma$} \label{mmbulgemsigma}

To estimate M$_{BH}$, we use the \cite{McConnell:2013} and \cite{Kormendy2013} power law scaling relations for M$_{BH}-$M$_{\mathrm{bulge}}$ and M$_{BH}-\sigma$ with intercept $\alpha$ and slope $\beta$:

\begin{equation}
    log_{10}(M_{BH }/M_{\odot}) = \alpha + \beta\log_{10}X,
\label{eq4}
\end{equation}

Where X=$\sigma/200\mathrm{ \ km \ s}^{-1}$ or $M_{\mathrm{bulge}}/10^{11}M_{\odot}$.
For \cite{McConnell:2013} M$_{BH}-$M$_{\mathrm{bulge}}$, we use the best-fitting power law for all galaxies where $\alpha=8.46\pm0.08$ and $\beta=1.05\pm0.11$. For M$_{BH}-\sigma$, we use $\alpha=8.32\pm0.05$ and $\beta=5.64\pm0.32$. \cite{McConnell:2013} also provide two separate M$_{BH}-\sigma$ relations using only early- and late-type galaxies and a log-quadratic fit for M$_{BH}-\sigma$. We test these additional scaling relations on our sample and found that the difference between these results and results from the single, log-linear M$_{BH}-\sigma$ relation were not statistically significant.

For \cite{Kormendy2013}, we use M$_{BH}-$M$_{\mathrm{bulge}}$ where $\alpha=8.69\pm0.04$ and $\beta=1.16\pm0.06$ and M$_{BH}-\sigma$ where $\alpha=8.49\pm0.05$ and $\beta=4.38\pm0.29$. It is important to note that \cite{Kormendy2013} excludes late-type galaxies from their dataset when fitting scaling relations. 

We use Monte-Carlo error propagation for errors, representing the host galaxy property and associated uncertainty as a Gaussian distribution. We do this by taking 100 random draws from a normal distribution with the catalog value as the mean and the error as the standard deviation. For $M_{\mathrm{bulge}}$, we randomly sample from a two-sided Gaussian distribution to capture the asymmetric errorbars in \cite{mt14}. We then calculate the black hole mass distribution for each galaxy. We quote the median value as the black hole mass and the standard deviation as the uncertainty. We found that 100 draws is the optimal number for our sample.

\subsection{Single-epoch virial masses}
\begin{figure}
    \centering
    \includegraphics[width=0.5\textwidth]{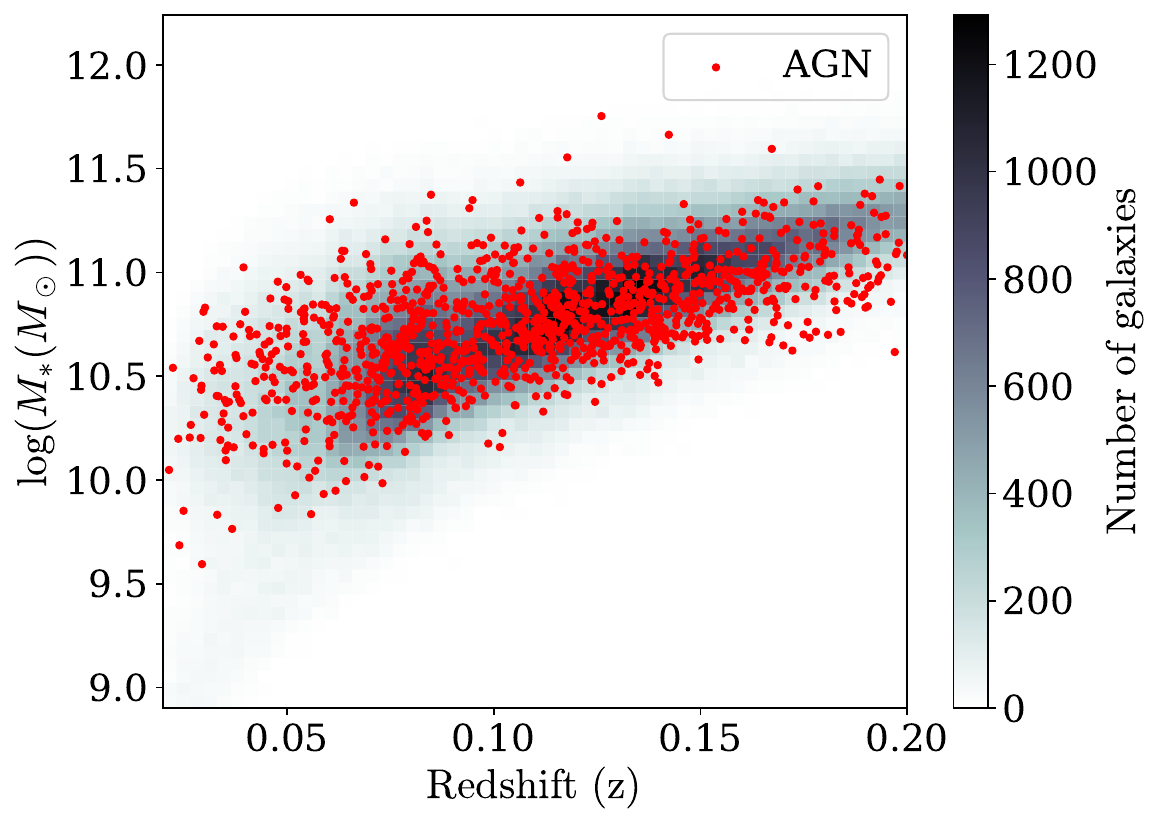}
    \caption{Here, our sample of 1,240 Type 1 AGN are red points overplotted on the entire sample of 413,538 galaxies on a total mass vs. redshift plot (same as in Figure \ref{fig:totalmassredshift}). \label{fig:totalmassredshiftagn}}
\end{figure}

We cross-match our sample with the \cite{Liu:2019} catalog of Type 1 broad-line AGN in SDSS and calculate single-epoch SMBH masses. In total, we have 1,240 Type 1 AGN in our sample. Figure \ref{fig:totalmassredshiftagn} shows their distribution in total mass and redshift. 

Type 1 AGN hosts only cover a limited total stellar mass range, and we want to determine the total stellar mass range where the AGN hosts are representative of the entire sample. We used the Hellinger distance to measure the statisticial similarity between the total mass distributions, finding that the Hellinger distance calculated between the $\log(\mathrm{M}_*/\mathrm{M_\odot})$ distributions of the AGN hosts and the quiescent galaxies is 0.15 when we restricted the quiescent galaxy sample to $10 < \log(\mathrm{M}_*/\mathrm{M_\odot}) < 11.5$, which is close to 0, indicating that the distributions are statistically similar. We conclude that our results in Section \ref{singleepoch} are only representative of our sample of quiescent galaxies with $10 < \log(\mathrm{M}_*/\mathrm{M_\odot}) < 11.5$. 

For galaxies that host Type 1 AGN, SMBH mass can be measured from some indicator of the velocity dispersion ($\sigma$) of the broad-line region (BLR) and size ($R_{BLR}$), where 
\begin{equation}
    M_{BH} = f\frac{\sigma^2R_{BLR}}{G},
\end{equation}
and $f$ is the virial factor that encodes assumptions about the unknown geometry of the BLR.

$R_{BLR}$ is measured by reverberation mapping, where the light-travel time across the BLR corresponds to the time lag between continuum flux variability and the response in the broad-line flux \citep{Blandford1982}. Reverberation mapping establishes the BLR radius-luminosity relation \citep{Kaspi2000, Bentz2013}, which is used to get single-epoch (SE) virial black hole masses from the optical AGN luminosity and broad-line widths \citep[e.g.,][]{GreeneHo2005,VP2006}. 

\cite{Liu:2019} reports single-epoch black hole masses using the relation from \cite{HoKim2015} for broad-line H$\beta$, but the virial factor from the reverberation-mapped black hole masses is obtained by assuming the $M_{BH} - \sigma$ relation. Therefore, we need a new single-epoch mass estimator that bypasses $M_{BH} - \sigma$ to avoid bias in our comparison to scaling relations.

The Sloan Digital Sky Survey Reverberation Mapping (SDSS-RM) project \citep{Shen2024} fits new single-epoch mass estimators to RM masses with an average virial factor $f$ obtained by comparing the virial product to the dynamical SMBH mass of $\sim$30 RM AGN. They get an average logarithmic viral factor $\langle\log f\rangle = 0.62 \pm0.07$. Although this is consistent with $\langle\log f\rangle$ determined from assuming scaling relations, this avoids a circular comparison with $M_{BH} - \sigma$ when comparing our results to single-epoch SMBH masses. However, we stress that these masses cannot be treated as ground truth for testing scaling relation accuracy due to assumptions about BLR modeling and geometry encoded in the virial factor.

We calculate $M_{SE}$ from the SE mass recipe in \cite{Shen2024} based on broad-line H$\beta$ where

\begin{multline}
\label{eq8}
\log \left( \frac{M_{\mathrm{SE,H}\beta}}{M_{\odot}} \right) = 
\\
\log \left[\left(\frac{L_{5100,\mathrm{AGN}}}{10^{44}\mathrm{ \ erg \ s}^{-1}}\right)^{0.5}\left(\frac{\mathrm{FWHM}}{\mathrm{km \ s}^{-1}}\right)^2\right]
+0.85,
\end{multline}
with an intrinsic scatter of 0.45 dex, where $L_{5100}$ is the monochromatic luminosity at $5100$\r{A} in $\mathrm{ \ erg \ s}^{-1}$ and the FWHM is the full-width half maximum of broad-line H$\beta$ in $\mathrm{km \ s}^{-1}$.
\cite{Shen2024} also give SE mass recipes using broad-line H$\alpha$, MgII, and CIV. We use the H$\beta$ estimator for our analysis since \cite{Shen2024} show it correlates best with the SMBH masses from reverberation mapping.

\section{Results} \label{results}

\subsection{M$_{BH}-$M$_{\mathrm{bulge}}$ vs. M$_{BH}-\sigma$}

\begin{figure*}[ht]
\gridline{\fig{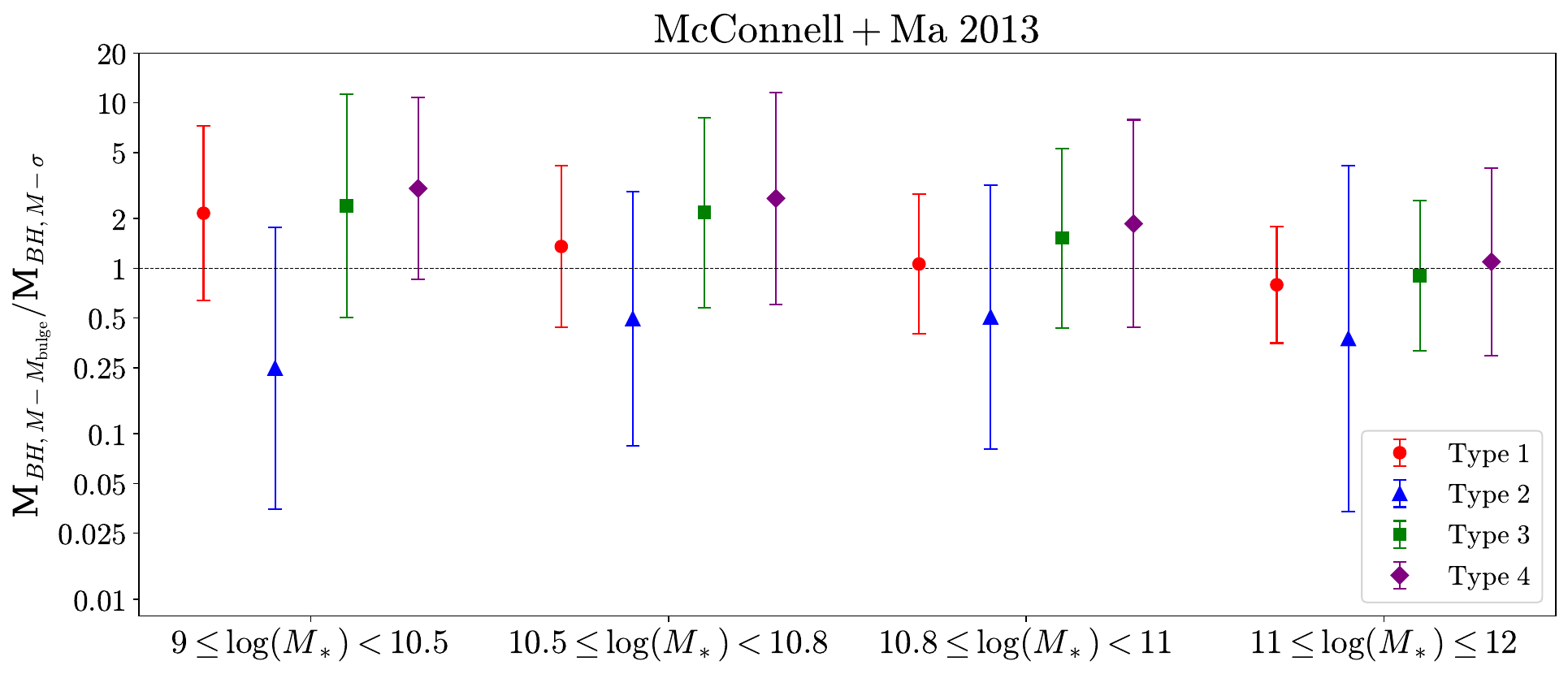}{0.8\textwidth}{(a)}}
\gridline{\fig{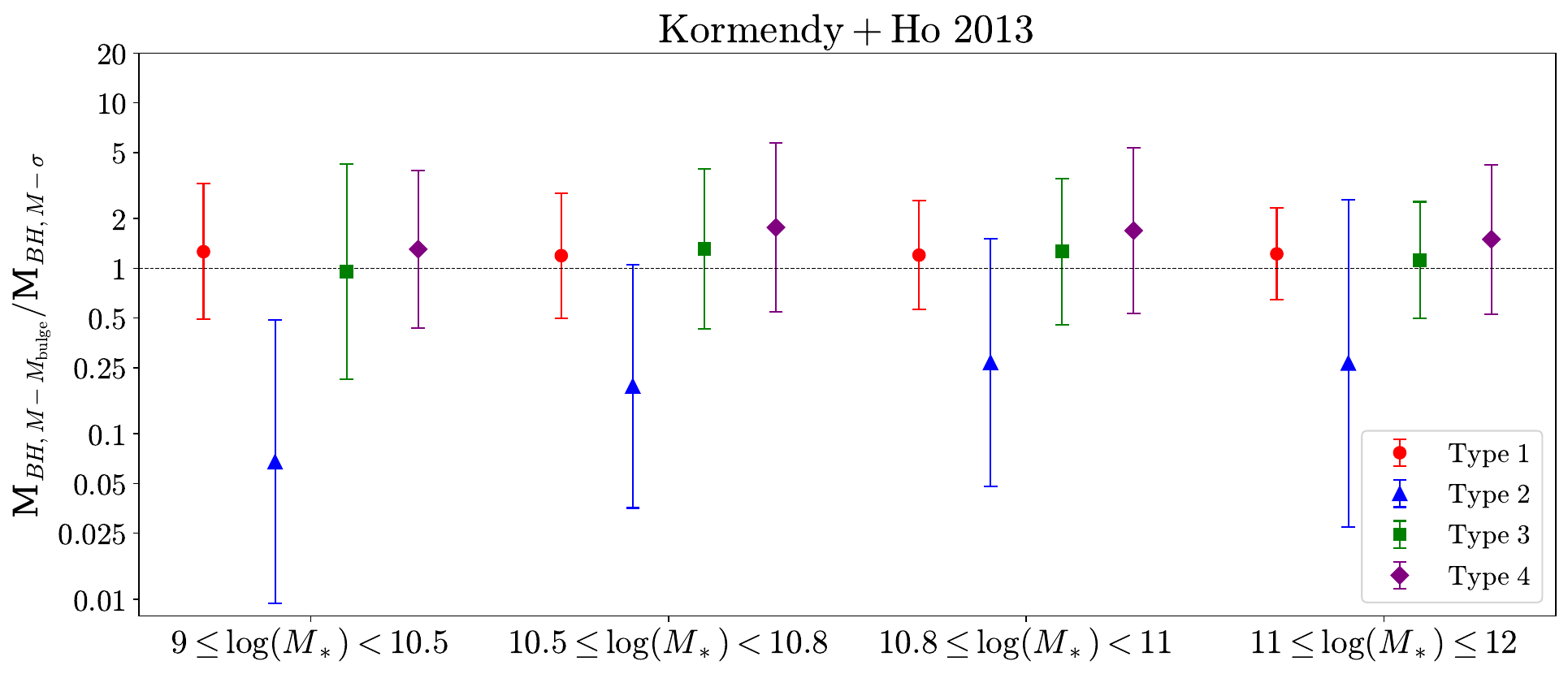}{0.8\textwidth}{(b)}}
    \caption{The SMBH mass ratio for total mass and galaxy type bins, where the types are spaced out on the x-axis for clarity. Type 1 galaxies are single-component bulges, Type 2 are disk-dominated galaxies, Type 3 are bulge+disk systems, and Type 4 are irregular galaxies. Figure (a) shows the offset between the \cite{McConnell:2013} scaling relations, where M$_{BH}-$M$_{\mathrm{bulge}}$ predicts more massive SMBHs than M$_{BH}-\sigma$ for low mass Type 1, 3, and 4 galaxies. For Type 2 galaxies, M$_{BH}-$M$_{\mathrm{bulge}}$ predicts less massive SMBHs for all total stellar masses. Figure (b) shows the scaling relations from \cite{Kormendy2013}, which has more agreement at lower total masses for Types 1, 3, and 4 than \cite{McConnell:2013} and a more extreme offset towards less massive SMBHs predicted by M$_{BH}-$M$_{\mathrm{bulge}}$ for Type 2 galaxies in the lowest total mass bin.  \label{fig:totalmasstype} }
\end{figure*}

To compare M$_{BH}-$M$_{\mathrm{bulge}}$ to M$_{BH}-\sigma$, we calculate the ratio M$_{BH,M-M_{\mathrm{bulge}}}/$M$_{BH,M-\sigma}$ for each galaxy in our sample. In log-space, the mass ratio distributions are roughly Gaussian, and indicate the direction which M$_{BH}-$M$_{\mathrm{bulge}}$ and M$_{BH}-\sigma$ are offset relative to each other.

To explore how the ratio of median SMBH mass depends on galaxy total stellar mass and morphology, Figure \ref{fig:totalmasstype} shows the sample binned by total stellar mass and galaxy type from \cite{mt14}. For the \cite{McConnell:2013} scaling relations, Type 1 (bulges), 3 (bulge+disk), and 4 (irregular) galaxies follow the same trend, with the offset between  M$_{BH}-$M$_{\mathrm{bulge}}$ and M$_{BH}-\sigma$ decreasing with increasing total mass. The spread about the median offset for these galaxy types at all total mass ranges is around 0.5 dex. For the Type 2 (disk) galaxies,  M$_{BH}-$M$_{\mathrm{bulge}}$ consistently predicts lower mass SMBHs than M$_{BH}-\sigma$ with $\sim1$ dex spread about the median offset. For the \cite{Kormendy2013} scaling relations, The Type 1, 3, and 4 galaxies show better agreement between M$_{BH}-$M$_{\mathrm{bulge}}$ and M$_{BH}-\sigma$ with $\sim0.3-0.5$ dex spread about the median. This suggests that fitting M$_{BH}-\sigma$ to early-type galaxies only will result in higher mass SMBH predictions, bringing these masses into agreement with those from M$_{BH}-$M$_{\mathrm{bulge}}$. For the Type 2 galaxies, \cite{Kormendy2013} shows the same results as \cite{McConnell:2013} where  M$_{BH}-$M$_{\mathrm{bulge}}$ predicts less massive SMBHs than M$_{BH}-\sigma$, with an even greater divergence in M$_{BH}-$M$_{\mathrm{bulge}}$ and M$_{BH}-\sigma$ in the lowest total stellar mass bin. 

Figure \ref{fig:totalmasstype} indicates consistency between M$_{BH}-$M$_{\mathrm{bulge}}$ and M$_{BH}-\sigma$ for galaxies with a prominent bulge component and high total stellar mass, reflecting the galaxy types accounted for in both M$_{BH}-$M$_{\mathrm{bulge}}$ and M$_{BH}-\sigma$. However, scaling relations give inconsistent SMBH mass predictions when applied to other galaxies beyond the subset of high-mass galaxies with prominent bulges. M$_{BH}-$M$_{\mathrm{bulge}}$ predicts much less massive SMBHs for galaxies with a pseudobulge regardless of the total galaxy mass. Even for galaxies with a more substantial bulge component, the lower mass bins show surprisingly large divergence between M$_{BH}-$M$_{\mathrm{bulge}}$ and M$_{BH}-\sigma$ where M$_{BH}-$M$_{\mathrm{bulge}}$ from \cite{McConnell:2013} tends to predict higher SMBH masses than M$_{BH}-\sigma$. Therefore, the choice of scaling relation significantly impacts the inferred SMBH population for galaxies with lower total stellar mass or prominent disks.

The divergence and spread in black hole masses could be due to the non-log-linear M$_{\mathrm{sph}}-\sigma$ relation with a bend at M$_{\mathrm{sph}} \sim 10^{10}$ M$_\odot$ \citep{c2013}, combined with the inherent bias in the galaxy samples used to fit  M$_{BH}-$M$_{\mathrm{bulge}}$ and M$_{BH}-\sigma$. It is unfeasible to individually resolve the spheroid component of low-mass spiral galaxies with low-quality observations, which prevents our ability to correctly apply a non-log-linear scaling relation.

\subsection{M$_{BH}-$M$_{\mathrm{bulge}}/\sigma$ vs. M$_{BH}-\mathrm{single \ epoch \ } H\beta $}\label{singleepoch}

\begin{figure*}[ht]
\gridline{\fig{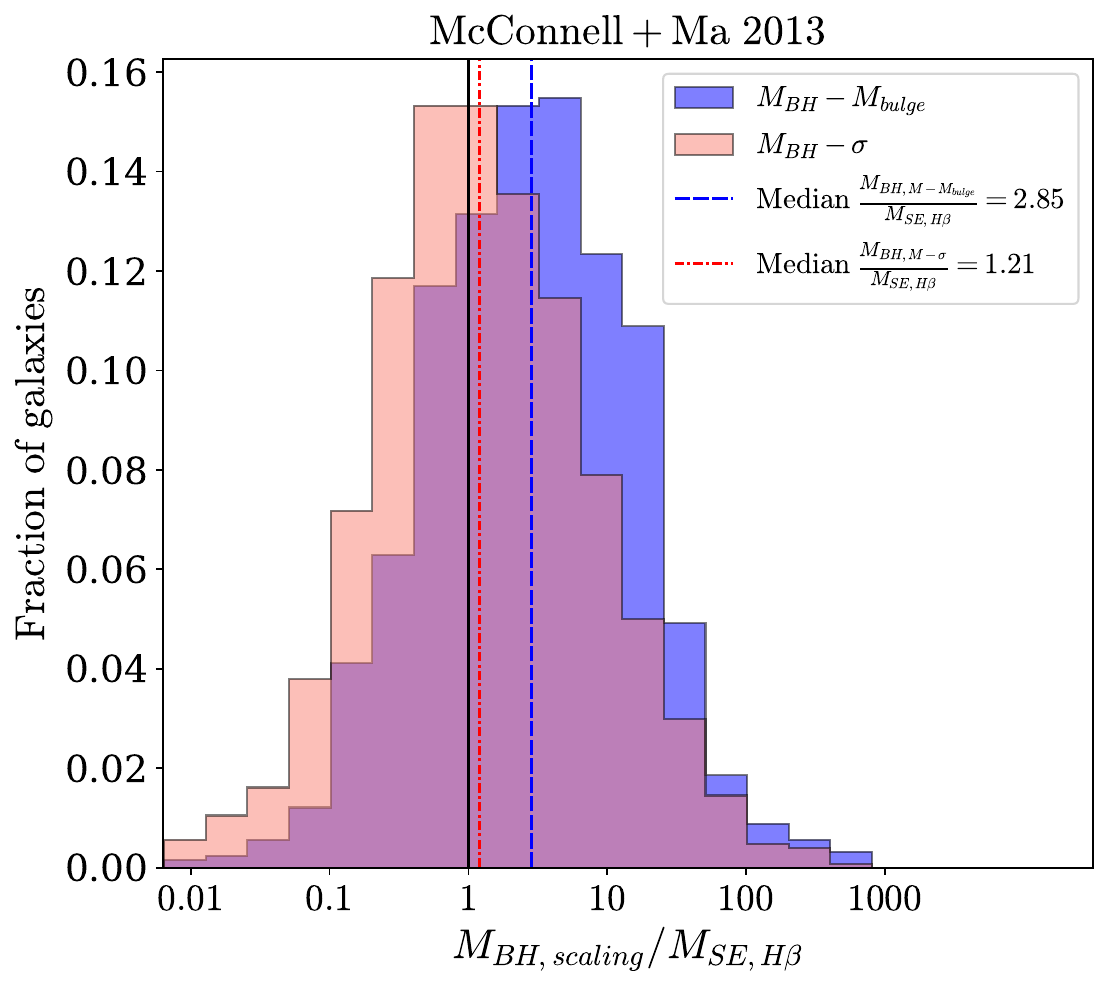}{0.5\textwidth}{(a)}
          \fig{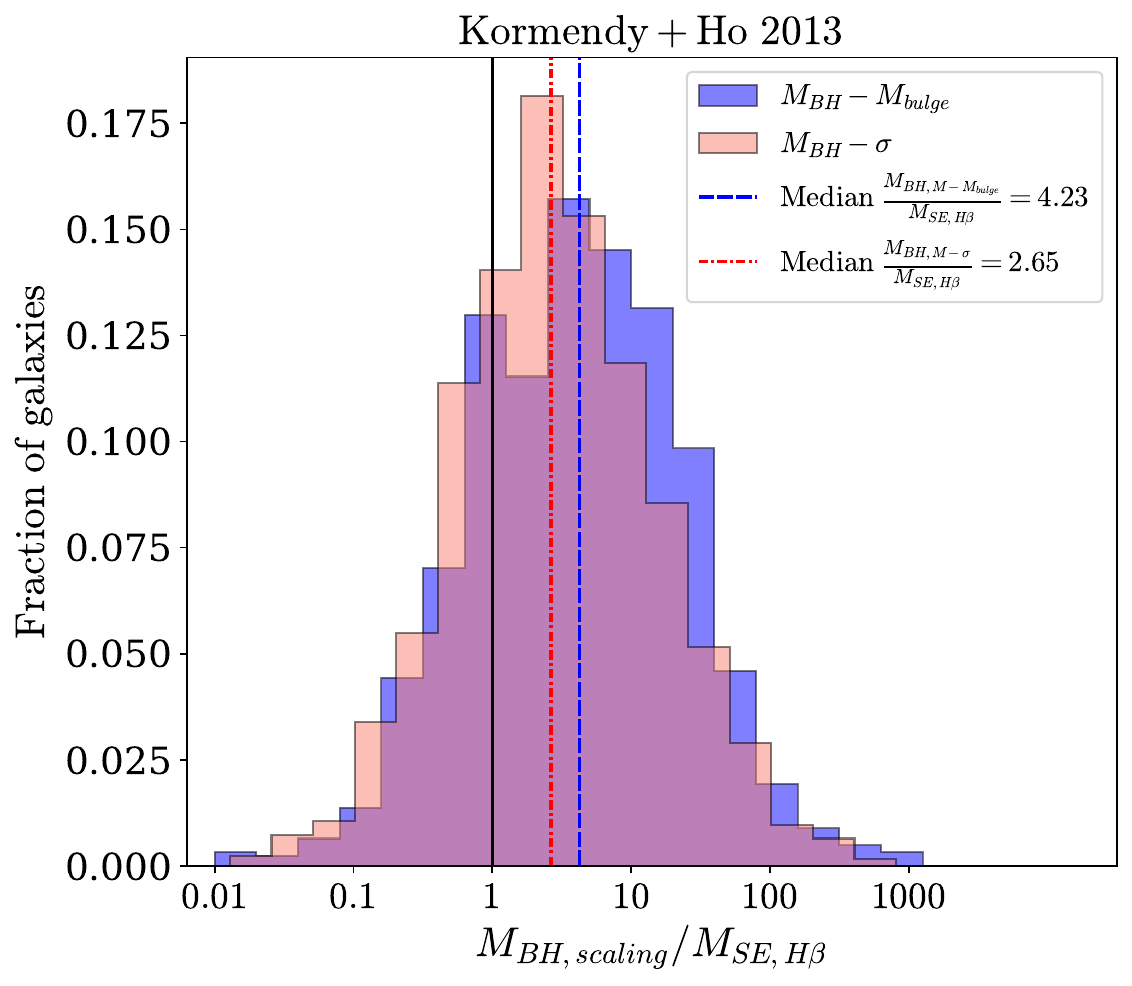}{0.5\textwidth}{(b)}}
    \caption{Normalized histogram of the ratio between the SMBH mass scaling relations (Equation \ref{eq4}) and the single-epoch mass estimator from H$\beta$ (Equation \ref{eq8}) on a log scale. The red and blue lines represent the medians of the mass ratio with M$_{BH} - \sigma$ and M$_{BH} -$M$_{\mathrm{bulge}}$ respectively. The black line at 1 represents perfect agreement between scaling relations and single-epoch mass estimators. For Figure (a), M$_{BH} - \sigma$ is closer to the one-to-one line, while the median of M$_{BH} -$M$_{\mathrm{bulge}}$ is offset by a factor of 3. For Figure (b), both M$_{BH} - \sigma$ and M$_{BH} -$M$_{\mathrm{bulge}}$ are offset toward higher masses. M$_{BH} - \sigma$ is offset by 3, and M$_{BH} -$M$_{\mathrm{bulge}}$ is offset by 4. \label{fig:singleepoch}}
\end{figure*}

\begin{figure*}[ht]
\gridline{\fig{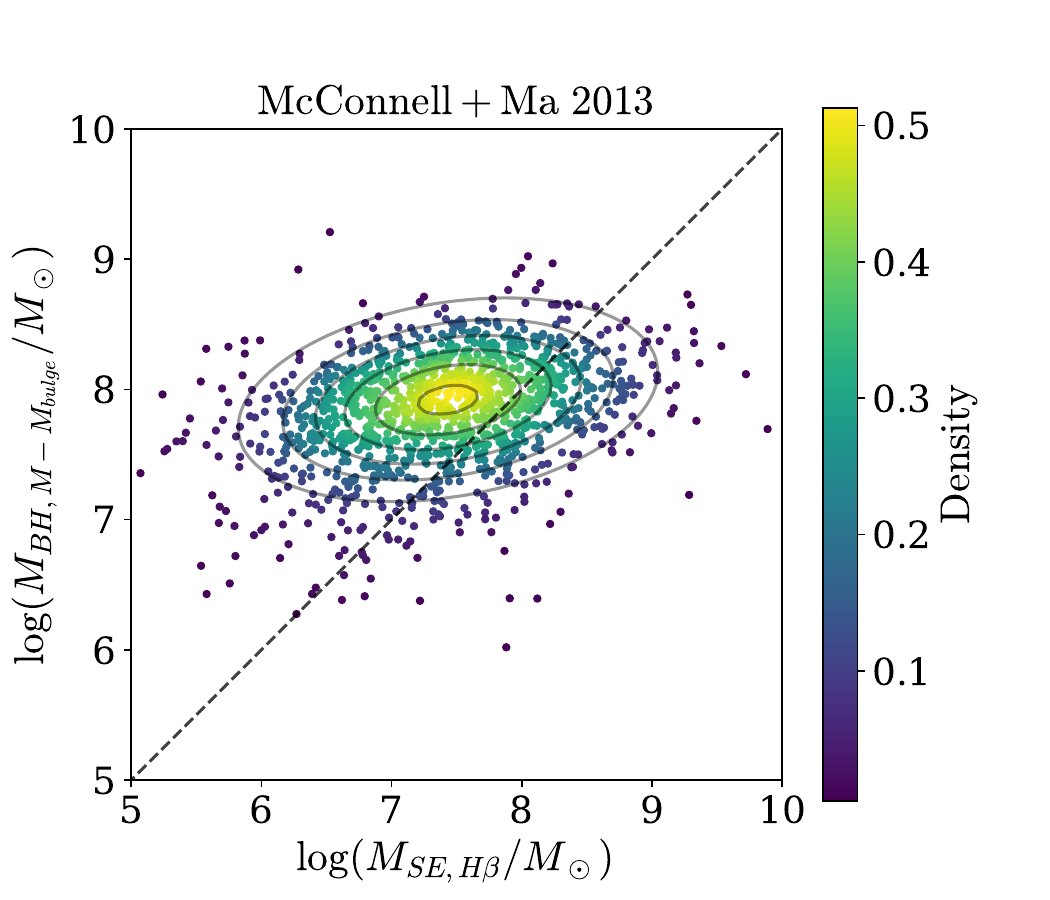}{0.5\textwidth}{(a)}
          \fig{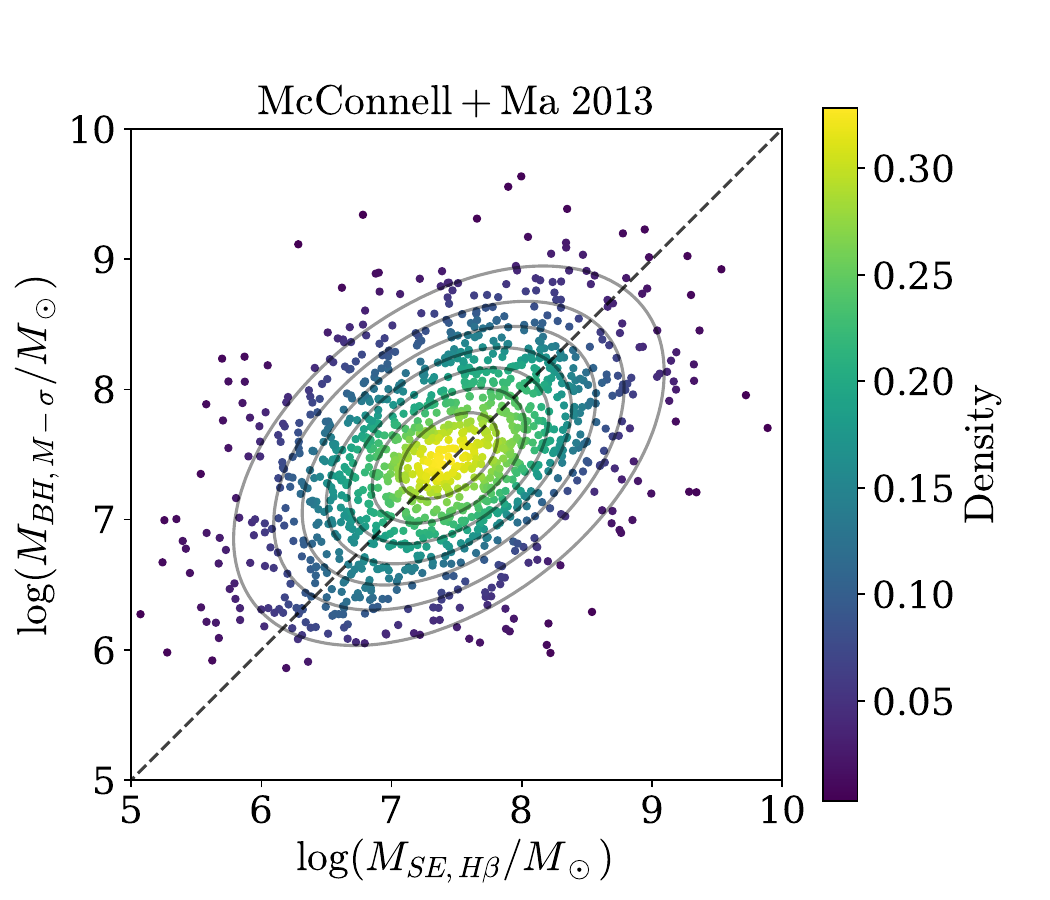}{0.5\textwidth}{(b)}
          }
\gridline{\fig{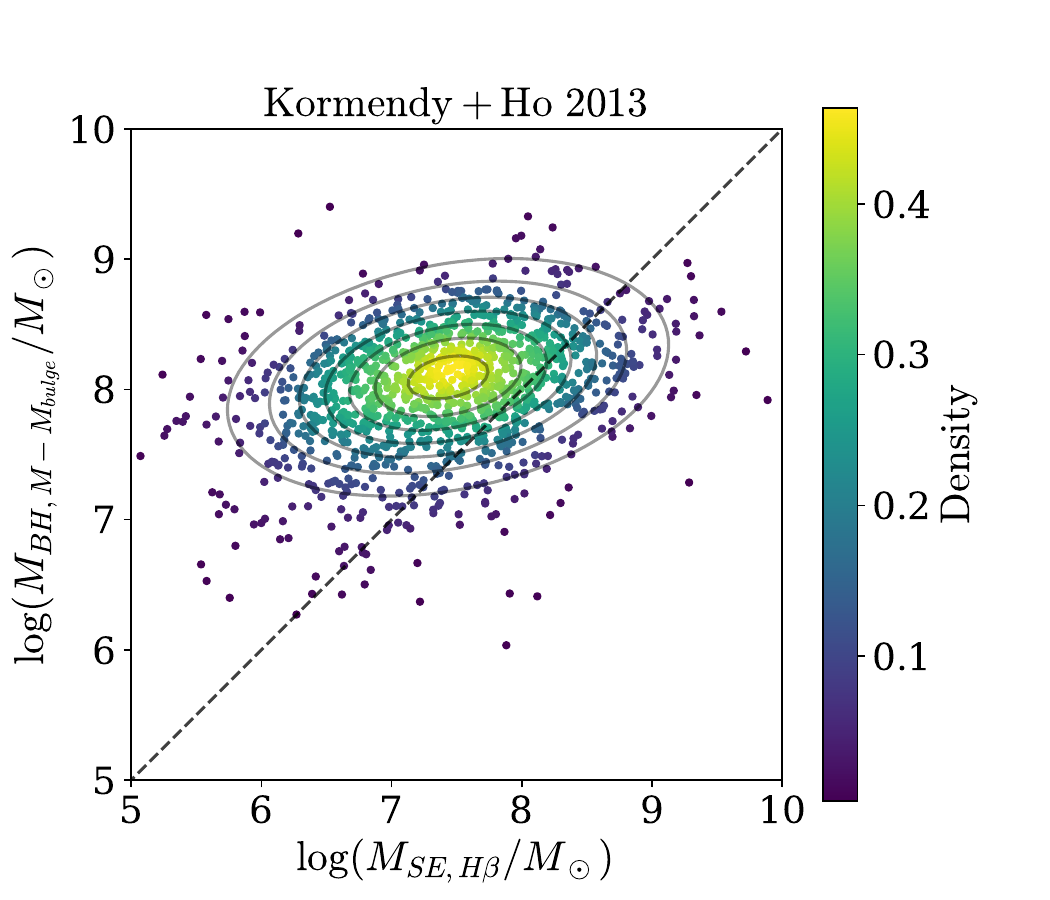}{0.5\textwidth}{(c)}
          \fig{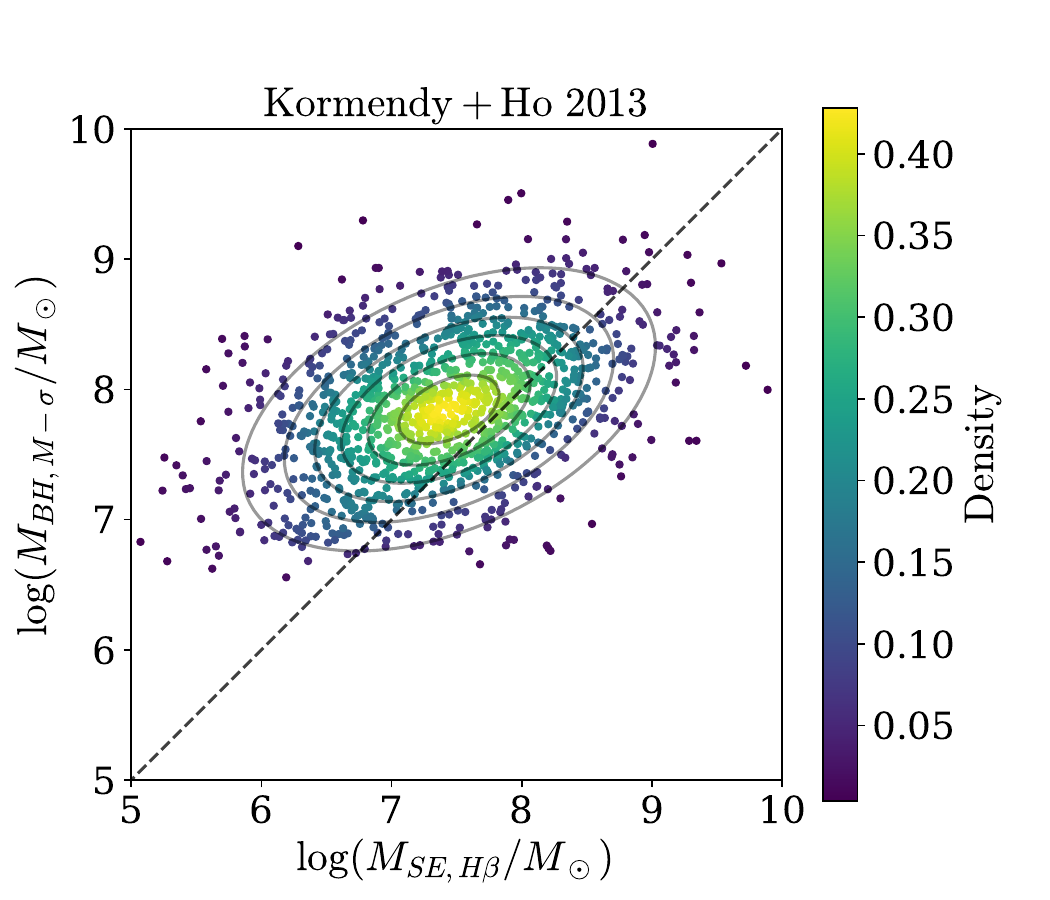}{0.5\textwidth}{(d)}
          }
    \caption{Scatter plot showing the SMBH mass from scaling relations (Equation \ref{eq4}) vs. the SMBH mass from the single-epoch mass estimator from H$\beta$ (Equation \ref{eq8}) on a log scale. The contours are the 2D Gaussian models with parameters in Table \ref{tab:2Dgaussian}. Figure (a) has the SMBH mass from the $M_{BH}-M_{\mathrm{bulge}}$ relation on the y-axis and Figure (b) has the SMBH from the $M_{BH}-\sigma$ relation on the y-axis. The colorbar is the Gaussian kernel density estimator that shows the concentration of points and the dashed line indicates a one-to-one linear relationship. Figure (a) shows that using $M_{BH}-M_{\mathrm{bulge}}$ predicts higher mass black holes than the single-epoch estimator, and Figure (b) shows that the single-epoch SMBH masses agree more with $M_{BH}-\sigma$ as it is centered on the one-to-one line. Figures (c) and (d) show that both scaling relations from \cite{Kormendy2013} are offset toward higher SMBH masses. \label{fig:2Dsingleepoch}}
\end{figure*}

\begin{deluxetable*}{ccccccc}
\tablecaption{2D Gaussian fits to $\log(M_{BH,\mathrm{scaling \ relation}})$ vs. $\log(M_{SE,H\beta})$ 
 \label{tab:2Dgaussian}}
\tablehead{\colhead{Scaling relation} & \colhead{Amplitude} & \colhead{$\mu_{x}$} & \colhead{$\mu_{y}$} & \colhead{$\sigma_{x}$} & \colhead{$\sigma_{y}$} & \colhead{$\theta$}}
\startdata
McConnell+Ma 2013 $M_{BH}-\sigma$            & 0.30 & 7.45 & 7.53 & 0.98 & 0.58 & 44.59 \\
McConnell+Ma 2013 $M_{BH}-M_{\mathrm{bulge}}$ & 0.52 & 7.44 & 7.92 & 0.85 & 0.37 & 44.18 \\
Kormendy+Ho 2013 $M_{BH}-\sigma$                            & 0.42 & 7.41 & 7.87 & 0.83 & 0.45 & 44.43 \\
Kormendy+Ho 2013 $M_{BH}-M_{\mathrm{bulge}}$                 & 0.48 & 7.42 & 8.10 & 0.82 & 0.40 & 44.25
\enddata
\end{deluxetable*}

To observe how M$_{BH}-$M$_{\mathrm{bulge}}$ and M$_{BH}-\sigma$ compare to other SMBH mass estimators, we calculate M$_{SE,H\beta}$ from \cite{Shen2024} (Equation \ref{eq8}) for the 1,240 Type 1 AGN in our sample. This mass estimator avoids assuming the M$_{BH}-\sigma$ relation when fitting for the virial factor. Still, there are other important assumptions to be cautious of when interpreting results from single-epoch SMBH masses. The viral factor still encodes assumptions about the unknown geometry of the BLR, and we assume active and quiescent galaxies follow the same empirical SMBH mass scaling relations \citep[e.g.,][]{Nelson2004,Woo2013,Caglar2020}.  

With these assumptions in mind, we show the results of our comparison in Figures \ref{fig:singleepoch} and \ref{fig:2Dsingleepoch}. The offset for both M$_{BH}-$M$_{\mathrm{bulge}}$ and M$_{BH}-\sigma$ show similar scatter about the median. In Figure \ref{fig:singleepoch} (a), the \cite{McConnell:2013} M$_{BH}-$M$_{\mathrm{bulge}}$ estimates SMBHs $\sim$3$\times$ more massive than M$_{SE,H\beta}$ and M$_{BH}-\sigma$ estimates SMBHs $\sim$1.2$\times$ more massive than M$_{SE,H\beta}$. Figure \ref{fig:singleepoch} (b) shows that M$_{BH}-$M$_{\mathrm{bulge}}$ from \cite{Kormendy2013} predicts $\approx4\times$ more massive SMBHs than the single epoch masses, and M$_{BH}-\sigma$ predicts $\approx3\times$ more massive SMBHs. 

Figure \ref{fig:2Dsingleepoch} shows the same results as a scatter plot, fit with 2D Gaussians whose parameters are detailed in Table \ref{tab:2Dgaussian}. We find that the best agreement with the single-epoch SMBH masses is M$_{BH}-\sigma$ from \cite{McConnell:2013}. Although the fits from \cite{Kormendy2013} show better agreement between M$_{BH}-$M$_{\mathrm{bulge}}$ and M$_{BH}-\sigma$ in Figure \ref{fig:totalmasstype}, masses from M$_{BH}-\sigma$ seem to be brought out of agreement with the single-epoch SMBH masses when only early-type galaxies are included in the fit.

\cite{Sturm2024} made a similar comparison between M$_{BH}-$M$_{\mathrm{bulge}}$ from \cite{Kormendy2013} and the single-epoch mass estimator M$_{SE,H\alpha}$ from \cite{Reines2013}, finding an order of magnitude offset between M$_{BH}-$M$_{\mathrm{bulge}}$ and M$_{SE,H\alpha}$. They show that the M$_{\mathrm{bulge}}$ measurement is not biased by any discernible galaxy property. They also conclude that the AGN luminosity is not a significant contributor to the bulge mass, and their results are the same whether AGN luminosity is subtracted from the bulge or not. We observe a offset in M$_{BH}-$M$_{\mathrm{bulge}}$ toward higher mass SMBHs with a similar range of galaxy total mass, but the offset is less than an order of magnitude.

\begin{figure*}[ht]
\gridline{\fig{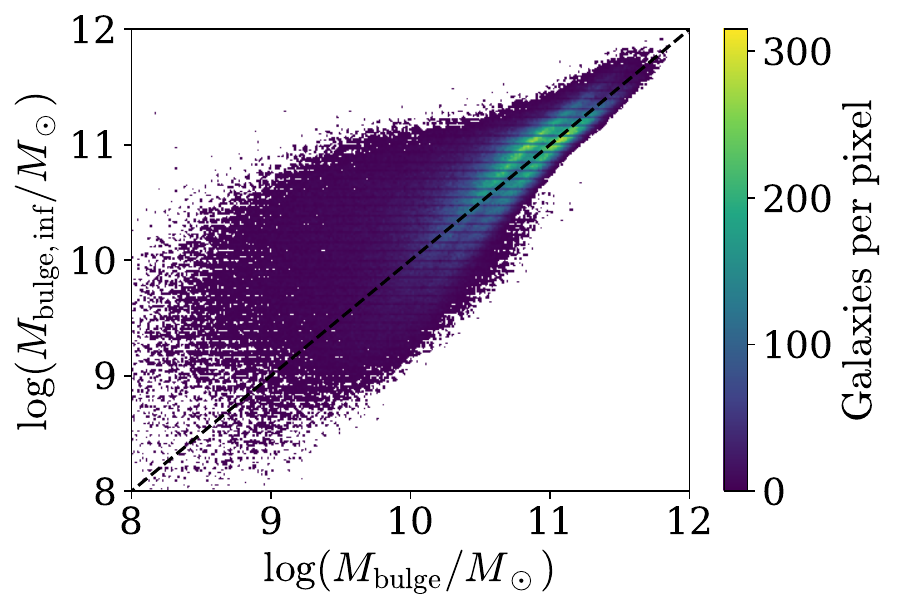}{0.5\textwidth}{(a)}
          \fig{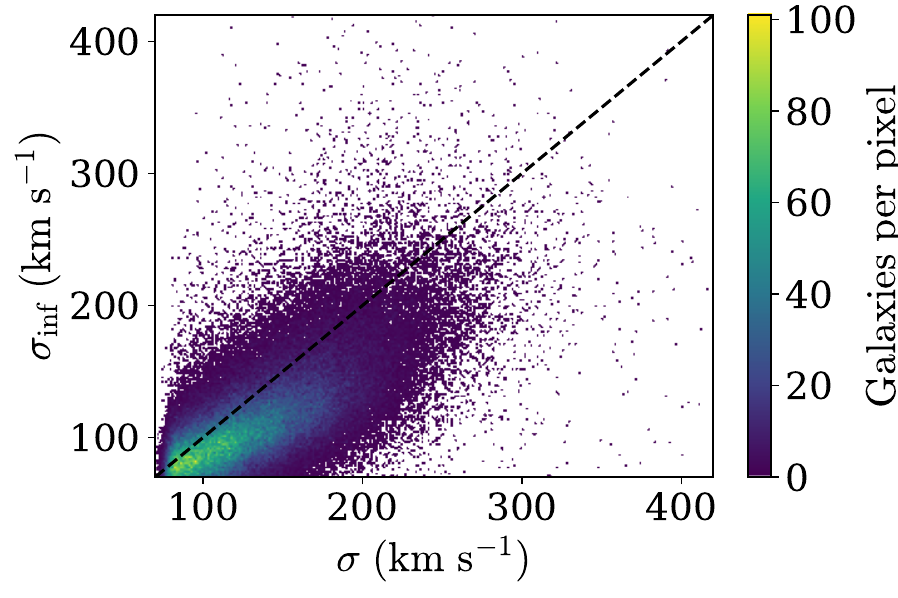}{0.5\textwidth}{(b)}
          }
\gridline{\fig{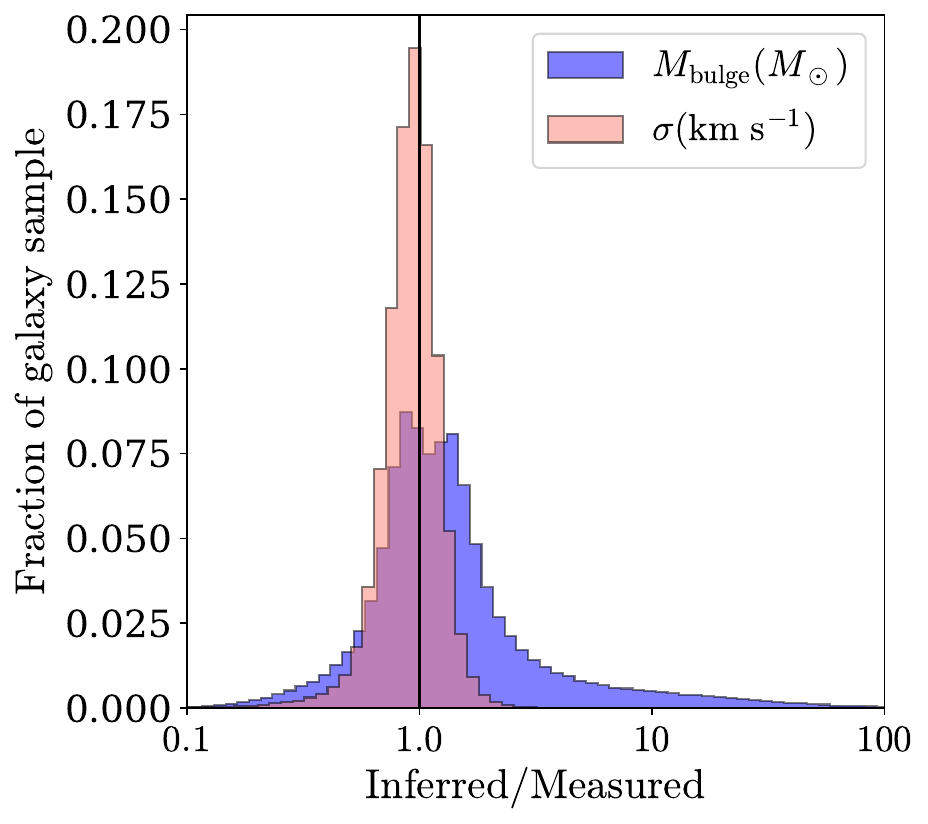}{0.5\textwidth}{(c)}}   
    \caption{Scatter plot showing the offset between inferred vs. directly measured galaxy properties with 1,240 AGN in each panel. Figure (a) has the $M_{\mathrm{bulge}}$ from bulge-disk decomposition on the x-axis and $M_{\mathrm{bulge}}$ inferred from total stellar mass on the y-axis. Figure (b) has $\sigma$ measured from SDSS spectroscopy on the x-axis and $\sigma_{inf}$ inferred from the virial theorem on the y-axis. The colorbar shows the concentration of points and the dashed line indicates a one-to-one linear relationship. Figure (a) shows outliers above the one-to-one line, where the inferred $M_{\mathrm{bulge}}$ is over-predicting compared to the bulge-disk decomposition. Figure (b) shows that the points are centered on the one-to-one line, indicating that inferred $\sigma$ is not biased in a particular direction compared to spectroscopy. A corresponding 1D histogram in Figure (c) compares the offset between inferred and measured galaxy properties. \label{fig:2Dinfvmeasured}}
\end{figure*}

\subsection{Inferred vs. measured host galaxy properties}\label{ivm}
When determining the GW signal amplitude from observational properties of SMBHs in the local universe, the black hole masses are calculated using scaling relations and basic photometry \citep[e.g.,][]{Sesana:2013,Ravi:2015,Arzoumanian:2021,astrointerp}. For velocity dispersion, the measured values correspond to aperture-corrected (Equation \ref{apcorr}) spectroscopic values from SDSS and the inferred values are from Equation \ref{eq3}.  We compared the inferred $f_{\mathrm{bulge}}$ to the bulge fraction measured from decomposition to determine the most accurate way of inferring $M_{\mathrm{bulge}}$. Figure \ref{fig:2Dinfvmeasured} shows the comparison between host galaxy properties inferred from photometry and measured directly.

The inferred velocity dispersion matches well with the velocity dispersion measured from spectroscopy, with the ratio between the two centered on one with little scatter about the median. For the inferred bulge mass, the ratio is also centered on one but there are large tails out to high inferred vs. measured bulge masses. This shows that while photometry can accurately infer velocity dispersion, this inference is more arbitrary for bulge mass. Inferred velocity dispersion includes more information about the galaxy overall, using morphology, size, and total mass. Inferred bulge mass is an educated guess on the bulge fraction based on broad assumptions of how the bulge fraction behaves with galaxy color and total mass. This most likely contributes to the high scatter between inferred and measured bulge mass, with inferred bulge fractions overestimating bulge mass by orders of magnitude in several cases.

\section{Conclusion} \label{conclusion}
In this study, we take advantage of the bulge-disk decompositions done on SDSS galaxies to create a large and well-defined, mass complete sample to test black hole-host galaxy scaling relations.  We calculate M$_{BH}-$M$_{\mathrm{bulge}}$ and M$_{BH}-\sigma$ using \cite{McConnell:2013} and \cite{Kormendy2013}, the single-epoch virial mass for Type 1 AGN M$_{SE,H\beta}$ using \cite{Shen2024}, and inferred M$_{\mathrm{bulge}}$ and $\sigma$ with \cite{Sesana:2013} and \cite{Bezanson:2011} respectively.

Our results reveal multiple limitations of SMBH mass scaling relations, such as the following:
\begin{itemize}
    \item We find that M$_{BH}-$M$_{\mathrm{bulge}}$ and M$_{BH}-\sigma$ are only consistent for high total mass ($\approx10^{11}-10^{12} M_\odot$) galaxies with a prominent bulge component. For disk-dominated galaxies hosting a weak pseudobulge (Type 2 in \citealt{mt14}), M$_{BH, M-M_{\mathrm{bulge}}}<$ M$_{BH, M-\sigma}$ by $\approx 3\times$ on average for all total stellar mass ranges. For galaxies with a substantial bulge component (Types 1 and 3 in \citealt{mt14}), \cite{McConnell:2013} shows M$_{BH, M-M_{\mathrm{bulge}}}>$ M$_{BH, M-\sigma}$ for $9 \leq \log(M_*/M_\odot) < 11$ with the offset between SMBH masses decreasing as $\log(M_*/M_\odot)$ increases. The offset decreases faster for Type 1 galaxies. Scaling relations from \cite{Kormendy2013} show agreement between M$_{BH}-$M$_{\mathrm{bulge}}$ and M$_{BH}-\sigma$ for Types 1, 3, and 4 in all total stellar mass bins. For Type 2 morphologies, we see the same results as the \cite{McConnell:2013} scaling relations where M$_{BH, M-M_{\mathrm{bulge}}}<$ M$_{BH, M-\sigma}$.
    \item We compared both M$_{BH}-$M$_{\mathrm{bulge}}$ and M$_{BH}-\sigma$ to single-epoch virial masses from broad-line H$\beta$ for 1,240 Type 1 AGN hosts in our sample. We find that overall, \cite{McConnell:2013} M$_{BH}-\sigma$ matches the single-epoch SMBH masses better than M$_{BH}-$M$_{\mathrm{bulge}}$ and M$_{BH}-\sigma$ from \cite{Kormendy2013}, where M$_{BH, M-M_{\mathrm{bulge}}}$ is greater than M$_{SE,H\beta}$ by $\approx 3-4\times$ on average. Due to the assumptions about modeling of the BLR, we stress that the single-epoch SMBH masses do not serve as ground truth in this study.
    \item  Finally, we infer photometric M$_{\mathrm{bulge}}$ and $\sigma$ and compare them to directly measured galaxy properties. We find that we can infer $\sigma$ much better than M$_{\mathrm{bulge}}$, with inferred M$_{\mathrm{bulge}}$ overestimating the bulge mass by orders of magnitude for several galaxies.
\end{itemize}

Black hole-host galaxy scaling relations play a fundamental role in characterizing SMBHs in our universe. Analysis of the GWB amplitude, cosmological simulations, and black hole-host galaxy coevolution models rely on scaling relations to predict SMBH masses. It is important to be cautious when applying scaling relations to entire galaxy populations as they quickly diverge when used for galaxies outside the narrow subset for which they were fit. This is especially important to consider when comparing the results from the gravitational wave background to local galaxy populations. The observational inputs used to predict the amplitude of the GWB and interpret astrophysics from the PTA signal often include an SMBH mass function based on the M$_{BH}-$M$_{\mathrm{bulge}}$ relation \citep[e.g.,][]{Sesana:2013,Ravi:2015,Arzoumanian:2021,astrointerp}. This paper shows that M$_{BH}-$M$_{\mathrm{bulge}}$ is consistent with M$_{BH}-\sigma$ at low redshift, but only for the subset of early-type (Type 1) galaxies with $\log(\mathrm{M}_*/\mathrm{M}_\odot) \gtrsim 11$. The tension between the amplitude of the GWB and predictions based on M$_{BH}-$M$_{\mathrm{bulge}}$ could imply a hidden population of SMBHs in the local universe \citep{sp2024,sp2025}. However, using the tension between the GWB and astrophysical observations to make predictions about SMBH populations for local galaxies, which may include systems with a prominent disk component or low total stellar mass, is essentially extending scaling relations to a regime that they were not designed for. In this paper, we have shown that using one scaling relation over another can significantly change the black hole masses estimated for these galaxies.

Although we cannot generalize that one scaling relation is more accurate than another, our results indicate that M$_{BH}-\sigma$ is able to reproduce the SMBH mass predictions from broadline H$\beta$ more closely than M$_{BH}-$M$_{\mathrm{bulge}}$. This is in line with recent studies that suggest M$_{BH}-\sigma$ is a more fundamental relation between SMBHs and their host galaxies than M$_{BH}-$M$_{\mathrm{bulge}}$ \citep[e.g.,][]{wake2012,vandenbosch2016,denicola2019,marsden2020,shankar2025}. 

In the future, a ground truth SMBH mass estimate could untangle the accuracy of black hole mass scaling relations. Until then, it is necessary to understand the assumptions underlying methods of SMBH mass estimation and to which galaxy populations they can be effectively applied.
\section*{Acknowledgments}
M. C. H., J.S., and J.M.C. are supported by NSF AST-1847938.
J.S. is supported by an NSF Astronomy and Astrophysics Postdoctoral Fellowship under award AST-2202388.

Funding for the Sloan Digital Sky Survey (SDSS) has been provided by the Alfred P. Sloan Foundation, the Participating Institutions, the National Aeronautics and Space Administration, the National Science Foundation, the U.S. Department of Energy, the Japanese Monbukagakusho, and the Max Planck Society. The SDSS Web site is http://www.sdss.org/.

The SDSS is managed by the Astrophysical Research Consortium (ARC) for the Participating Institutions. The Participating Institutions are The University of Chicago, Fermilab, the Institute for Advanced Study, the Japan Participation Group, The Johns Hopkins University, Los Alamos National Laboratory, the Max-Planck-Institute for Astronomy (MPIA), the Max-Planck-Institute for Astrophysics (MPA), New Mexico State University, University of Pittsburgh, Princeton University, the United States Naval Observatory, and the University of Washington.

This work utilized the Alpine high performance computing resource at the University of Colorado Boulder. Alpine is jointly funded by the University of Colorado Boulder, the University of Colorado Anschutz, Colorado State University, and the National Science Foundation (award 2201538).

\bibliography{references}
\bibliographystyle{aasjournal}
\end{document}